\documentclass[prb,preprint,superscriptaddress,showkeys,floatfix]{revtex4}% Physical Review B PREPRINT

\usepackage{graphicx}% Include figure files
\usepackage{dcolumn}% Align table columns on decimal point
\usepackage{bm}% bold math

\newcommand{\bra}[1]{
\left<{#1}\right|
}
\newcommand{\ket}[1]{
\left|{#1}\right>
}

\begin{document}

\title{Real Space Multiple Scattering Calculations of Relativistic
Electron Energy Loss Spectra}% Force line breaks with \\

\author{K. Jorissen}
 \email{kevin.jorissen@ua.ac.be}
 \affiliation{Department of Physics, University of Washington, Seattle, WA
98195 U.S.A.}
 \affiliation{Universiteit Antwerpen, Groenenborgerlaan 171, B-2020 Antwerpen, Belgium}
\author{J.J. Rehr}%
 \affiliation{Department of Physics, University of Washington, Seattle, WA
98195 U.S.A.}

\date{\today}

\begin{abstract}
{\it Ab initio} calculations of relativistic electron energy loss spectra (REELS) are
carried out using a generalization of the real-space
Green's function code FEFF8 which is applicable to general aperiodic
materials.  Our approach incorporates relativistic effects
in terms of the cross-section tensor within the dipole selection rule.
In particular the approach explains relativistic corrections to the
magic angle in polarized EELS experiments.  Our generalization
includes instrumental effects, such as the integral of the cross
section over the impulse transfer dependence,
to account for a finite detector aperture and electron beam width.
The approach is illustrated with an application to the graphite C K edge. 
\end{abstract}

%Use showkeys class option if keyword display desired
\maketitle

\section{\label{intro}Introduction}

Electron energy loss spectroscopy (EELS) measures
the energy loss of a fine beam of high energy electrons
($\approx$ 100 keV) propagated through a sample
in an electron microscope.\cite{Egerton}  The energy loss spectrum
is defined as the fraction of electrons which has lost a given amount
of energy by interacting inelastically with the sample.
From the EELS spectrum, one can obtain structural, chemical and
electronic information similar to that in
x-ray absorption spectroscopy (XAS), encompassing both
extended x-ray absorption fine structure (EXAFS) and
x-ray absorption near edge structure (XANES).
We focus here on the ELNES (energy loss near edge structure)
edges in the spectrum, corresponding to inelastic losses through the
excitation of an electron from a deep core level into unoccupied states.  Because EELS is
an absorption technique, and the initial core level state is sharp in energy,
such core loss signals reflect the electronic 
structure of unoccupied electron states. In particular, apart from a smooth
background cross-section factor, the shape of the ionisation edge is roughly
an image of the unoccupied angular momentum projected density of electron
states (LDOS).
Due to selection rules, the observed spectra correspond to a
decomposition of the DOS according into various angular momentum
$l$ channels.
Also the possibility of tilting the specimen with respect to the electron
beam  at fixed scattering angle allows one to investigate the
anisotropy in the local unoccupied DOS because the scattering vector
$\textbf{q}$ (i.e., the momentum transfer) appears in the transition matrix
element.  Here ${\bf q} = {\bf k} - {\bf k'}$, where ${\bf k}$ is
the wave vector of the incident fast electron and ${\bf k'}$
the wave vector of the scattered fast electron.  This is the
analog of the linear dichroism well known in x-ray absorption
spectroscopy (XAS).

What distinguishes EELS from XAS and similar techniques is that one can
now obtain very local atomic scale information
by focussing a very small probe of width $\approx$ 0.1 nm  on a sample
in a transmission electron microscope (TEM).  In the age of nanotechnology,
this is relevant for studies of nanoscale materials.  Modern
instruments with field emitters also allow detection of ELNES with an energy
resolution of 0.6 to 0.7~eV on a sub-nanometer scale, and monochromated
transmission electron microscopes (TEM)
reach 0.1~eV resolution. \cite{Monochrom}

\textit{Ab initio} calculations of EELS have often been used to support
the interpretation of experimental data.  Several approaches for these
calculations have been developed. For example,
one approach for periodic structures is based on density functional theory (DFT)
and the LAPW band-structure code WIEN2K \cite{HebertM05,Wien2k} and super-cell
techniques.  An alternative approach for EELS makes use of the
real space Green's function based FEFF8 \cite{Ankudinov_feff8},
which is applicable to periodic and non-periodic structures alike.
Although this \textit{ab initio} code has been used extensively in the field of
x-ray spectroscopy, it has been frequently applied to EELS calculations
as well.  A recent review of the application of FEFF8 to EELS calculations
can be found in Ref.\ \onlinecite{Moreno}.  This approach is based on
the {\it approximate} equivalence between dipole-selected EELS (i.e., the
long wavelength limit) and XAS that has long been taken for granted. 
In this paper, however, we examine the quantitative differences between
XAS and EELS.  In particular we discuss the differences due to relativistic effects
as well as instrumental effects (e.g., 
characteristics of the electron microscope).  These are
needed to obtain quantitative agreement
with modern EELS experiments using relativistic beam energies.

Recently, it has been recognized \cite{SchattschneiderPRB05}
that a relativistic interaction Hamiltonian is essential for accurate
calculations of the scattering cross section, i.e., to 
obtain quantitative results for anisotropic materials at relativistic
beam energies.
The effect is a relativistic compression of the interaction field, which is therefore
anisotropic in the dipole limit. Thus, treating this relativistic
effect requires a generalization of the FEFF8 code
to account for the momentum-transfer dependence of relativistic
EELS experiments.
The main purpose of this work is to develop an approach for
\textit{ab initio} EELS calculations that builds in the relativistical formalism of Ref. \onlinecite{SchattschneiderPRB05}
within the framework of Green's function multiple-scattering theory.
In particular this effort builds on the self-consistent electronic-structure/spectroscopy code FEFF8,\cite{Ankudinov_feff8}
but introduces a number of extensions relevant to modern EELS experiments.
This development is broadly applicable and provides a new, general tool for
EELS calculations which is complementary to band-structure techniques
and is applicable over a broad spectral range.

One important application is the so-called {\it magic angle}, which
has been at the heart of the discussion in the literature on relativistic EELS.\cite{SchattschneiderPRB05}
This angle is defined as the value of the detector aperture
of an electron microscope for which the measured EELS
spectrum is independent of the relative orientation of sample and electron
beam.  This quantity is of direct practical importance for
polarized EELS experiments, in which the anisotropy of the signal
can be an unwelcome complication.  Notably the magic angle is
material independent, and therefore provides a direct test of the
validity of scattering theory that our
calculations are based on.  We show in this paper that 
relativistic calculations based on the new FEFF8 code
significantly improve on nonrelativistic calculations.

The theory relevant to the relativistic extension of FEFF8 is
briefly described in Sec. \ref{theory},
while Sec. \ref{pract} describes computational considerations.
Finally applications are given in Sec. \ref{app}.

\section{\label{theory}Theory}

The EELS signal can be described by the double differential scattering
cross section (DDCS)
\begin{equation}\label{ddscs}
{{\partial ^2 \sigma \left( {{\bf q},E} \right) } \over
{\partial \Omega \partial E}},
\end{equation}
which is the probability of detecting a scattered electron which has lost
energy $E$ and transferred-momentum ${\bf q}$ into the
solid-angle $d\Omega$.  Formally the DDCS can be expressed in terms of
the bare Thomson cross-section and the relativistic
dynamic structure factor $S({\bf q},E)$.
Since the Thomson cross-section is sharply peaked at small $q$, it
is common practise and generally a good approximation
to consider only the so-called dipole transitions (i.e., small $q$ limit)
where the orbital momentum quantum number $l$ of the atomic electron changes
by $\pm 1$ in transitions. 
Recently it has been shown \cite{SchattschneiderPRB05} that within the dipole approximation
the relativistic DDSCS for EELS is given by
\begin{eqnarray}
\frac{\partial ^2 \sigma \left( {{\bf q},E} \right) }
{\partial E\partial \Omega }& =&
\left( \frac{\partial \sigma } {\partial \Omega } \right)_{Th}
\text{S}( \bf{q'}, E )  \label{ddscs2} 
\end{eqnarray}
\begin{eqnarray} \label{cs}
\left( \frac{\partial \sigma } {\partial \Omega } \right)_{Th}& =& 
\frac{4a_0^{ - 2} \gamma^2 } {[q^2  - ( E / \hbar c)^2 ]^2}
\frac{k' } {k }
\end{eqnarray}
\begin{eqnarray} \label{dff}
\text{S}( {\bf q'}, E ) & =& \sum_{i,f} 
| \langle i | {\bf q'} \cdot {\bf r}
%(({\bf q}-\beta^2 q_z {\bf \hat e_z}).{\bf r}) 
|f \rangle  |^2 \delta (E_f - E_i  - E ). \label{crosssection}
\end{eqnarray}
Here the momentum transfer in the dipole transition element
is relativistically contracted, i.e.,
\begin{equation} \label{q'}
{\bf q'} = {\bf q}-\beta^2 q_z {\bf \hat{e_z}}
%\cdot {\bf r} ,
\end{equation}
where $\beta = v/c$ and $v$ is the beam velocity.
This equation is very similar to the description of XAS in the
dipole limit, with the impulse transfer ${\bf q}$ playing the role of
the polarization vector ${\bf \hat{\epsilon}}$
in x-ray scattering matrix elements.
However, for relativistic EELS there is an extra ${\bf q}$-dependent
contribution along  the direction of propagation ${\bf \hat{e_z}}$. 
\\ In general the DDCS can always be separated into a
probe-dependent part containing the ${\bf  q}$-dependence,
and a sample-dependent part which is independent of ${\bf q}$. Since
the theory is bi-linear in ${\bf q}$ the sample-dependent term transforms as
a tensor, i.e.,
\begin{equation} \label{sumsigma}
\text{S}({\bf q'},E) = \sum\limits_{i,j = 1}^3 {q'_i \,q'_j \,\sigma_{ij}(E)}
\label{sum} 
\end{equation},
\begin{equation}\label{sigma}
\sigma_{ij}(E)=\sum_{i,f}\bra{i}x_i\ket{f}\bra{f}x_j\ket{i}\delta(E-E_i+E_f)\;.
\end{equation}
The cross-section tensor $\sigma_{ij}$ (CST) can therefore describe all possible
transitions of the sample.  However, experimental conditions determine which
impulse transfers occur and therefore the weight of each component of the
cross-section tensor that contributes to the total cross section.  This can be
illustrated clearly by considering the sample to beam orientation of an EELS
experiment.  Rotation of the sample is equivalent to a rotation of ${\bf q}$,
thus changing the weights of the $\sigma_{ij}$ components in Eq.\
(\ref{sumsigma}).
The relativistic character of the formalism is also obvious: the field of the beam electron
contracts in its propagation direction, resulting in the evaluation of \ref{sumsigma}
using a contracted impulse transfer vector as in Eq.\ (\ref{q'}), denoted by a prime.

Formally the CST is a symmetric tensor with at most six independent components. 
As such, it can always be diagonalized.  However, only in symmetric materials,
where its principal axes are given by the physical symmetry of the crystal
itself, \textit{a priori} knowledge of the diagonal representation is available,
and one set of coordinates diagonalizes the tensor for all energies.
  In the general case of a low symmetry sample,
or in a situation where a non-symmetric coordinate system is desirable,
the cross terms in Eq.\ (\ref{sumsigma}) are important contributions to the
cross section which cannot be neglected.  We give an example of
this in Sec. \ref{app}.

\section{\label{pract} Practical considerations}

\subsection{\label{expt}Description of the experiment}
An \textit{ab initio} EELS calculation requires not only a description
of the sample
as input.  It also requires knowledge of the conditions in which the experiment was done.
We briefly discuss some experimental parameters specific to EELS.  An EELS experiment usually involves an integration over the DDCS defined in
Eq.\ (\ref{ddscs2}).
Typically, the probe has a certain angular width, characterized by the
convergence semi-angle $\alpha$, allowing a whole set of ${\bf k_i}$ 
incoming plane waves.  Similarly, the detector integrates the signal over a
certain range of outgoing beam directions, characterized by the collection
semi-angle $\beta$.  Both are usually of the order of mrad.  Assuming that
the incoming beam is monochromatic, the measured signal is then given by
\begin{equation}
{{\partial \sigma (E)} \over {\partial E}} = \int\limits_{\alpha ; \beta}  {{{\partial ^2 \sigma } \over {\partial \Omega \partial E}}\left( {{\bf q},E} \right)d^3q}
\label{dscs}
\end{equation}.
To cast the orientation dependence of the EELS spectrum in a more explicit form, we can rewrite Eq.\ (\ref{dscs})
using the CST :
\begin{equation}
{{\partial \sigma } \left(E \right) \over {\partial E}} = \sum\limits_{i,j = 1}^3 {\sigma _{ij}(E) \int\limits_{\alpha ; \beta}  {q'_i \,q'_j \ \frac{4a_0^{ - 2} \gamma^2} {[q^2  - ( E / \hbar c)^2 ]^2} \frac{k' } {k }   d^3q}}
\label{dscssum}
\end{equation}

As the CST depends only on the sample, only integrals of functions of
${\bf q}$ need to be calculated.
These integrals are approximated by a sum over a finite set of impulse transfer vectors ${\bf q}$. This is considered in more detail in the Appendix.

Other experimental parameters included in our calculations are the electron beam energy, the sample to beam orientation, and the position of the EELS detector in the scattering plane.  FEFF8 calculations always include core hole broadening.  Additional broadening can be applied.

\subsection{\label{comp} Computational details}

Our calculations are incorporated in a generalization of the
\textit{ab initio} real space multiple scattering code FEFF8, which we
call FEFF8.5.  We made modifications to FEFF8 to obtain the full-cross section
tensor $\sigma_{ij}$ from the program, and added a new module that
calculates the net cross section of Eq.\ (\ref{sumsigma}) for a given
${\bf q}$
from the cross section tensor, and carries out the integrals over ${\bf q}$
as described in Sec. \ref{expt}.  We have taken care to retain all other
features of the code, including the use of advanced
cards such as TDLDA in the calculations \cite{bsepaper} which account for
corrections to the independent electron approximation ; the use of 
Debye-Waller factors to approximately account for temperature effects ; etc.

The calculation of the cross section tensor for EELS is analogous to the case
of XAS calculations,\cite{Ankudinov_feff8}
\footnote{ See the FEFF documentation
http://leonardo.phys.washington.edu/feff/}.
  For the near edge region or energy loss near edge structure (ELNES)
- the electron equivalent of XANES), the full multiple scattering technique
(FMS) is used, in which all scattering paths within a
sphere of limited radius are summed implicitly by matrix inversion.  
For the extended region or extended energy loss fine structure 
(EXELFS) - the electron equivalent of EXAFS, the path expansion approach
is taken, in
which the scattering from a selected number of paths of limited length is
summed explicitly.  This is done for each of the six independent components of
the sigma tensor.  Combining the ELNES and EXELFS calculations, one can calculate
spectra over hundreds of eV, far beyond the limitations of most band-structure codes. 

Note that the calculation of the probe and the sample are treated separately
so that it is sufficient to calculate the properties of the sample
once for the simulation of many experimental situations.
We remark that our EELS code can also calculate mixed dynamic form factors
(MDFF), which are off diagonal in ${\bf q}$, but we postpone the discussion
of these to a future publication.

\section{\label{app} Applications}
\subsection{\label{graphite}The C K edge of graphite}
The strongly anisotropic nature of graphite makes its EELS spectra highly
susceptible to orientation effects and therefore to relativistic contributions to the cross section that we discussed in Sec. \ref{theory}.  We demonstrate
our method on the C K edge, which has an energy threshold of 285 eV.

\subsubsection{\label{CST}Cross-section tensor}
In this subsection, we discuss the components of the CST, and show that
calculation of its diagonal components is
not generally sufficient to calculate the EELS spectrum.  To see this,
we show the different components of the cross section tensor calculated
in two different coordinate systems.  System 1 is symmetrical : its $z$-axis is perpendicular to the graphene sheets of the sample, $x$ and y are in-plane.  System 2 is non-symmetrical : it is obtained from system 1 by a rotation of 35° around the $x$-axis of system 1.  We work in a Carthesian representation and refer to the components $i$,$j$ of $\sigma$ as $x$,$y$,$z$.
%xxx Fig 1

\begin{figure}
\includegraphics[width=0.3\textwidth,angle=270]{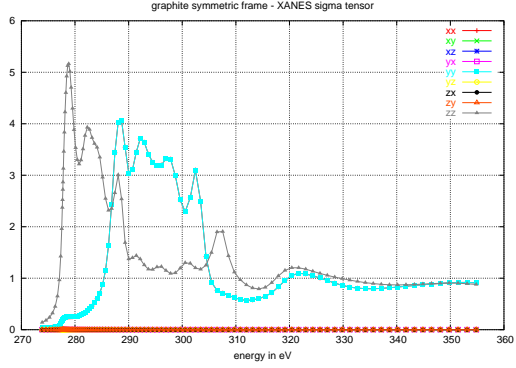}
\caption{Components of the CST of graphite in symmetric coordinates.}
\label{sigma_sym}
\end{figure}

\begin{figure}
\includegraphics[width=0.3\textwidth,angle=270]{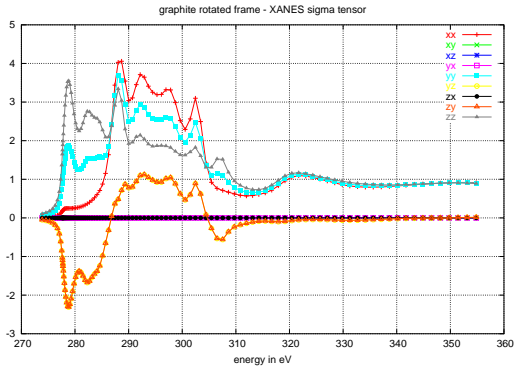}
\caption{Components of the CST of graphite in nonsymmetric coordinates, tilted 35° around the symmetric $x$-axis.}
\label{sigma_rot}
\end{figure}

In Fig.\ \ref{sigma_sym}, we see that in symmetric coordinates the $\sigma_{zz}$ spectrum contains the so-called $\pi$-transitions, while $\sigma_{xx}$ and $\sigma_{yy}$ are identical and contain the $\sigma$-transitions.  All off-diagonal components are zero, as can be explained by symmetry, i.e.,
equivalence of $x$ and -$x$, $y$ and -$y$, and $z$ and -$z$.

Fig.\ \ref{sigma_rot} shows that in the rotated system,
$\sigma_{xx}$ is equal to that in the symmetric frame, but $\sigma_{yy}$
and $\sigma_{zz}$ have mixed and are of mixed $\pi$, $\sigma$ - character. 
Additionally, the decrease in symmetry allows $y$,$z$ cross terms to exist. 
The $x$, -$x$ symmetry has been preserved, suppressing $xz$, $zx$, $xy$ and $yx$
components.  A more general rotation of the coordinates would make all
off-diagonal elements nonzero.

\begin{figure}
\includegraphics[width=0.3\textwidth,angle=270]{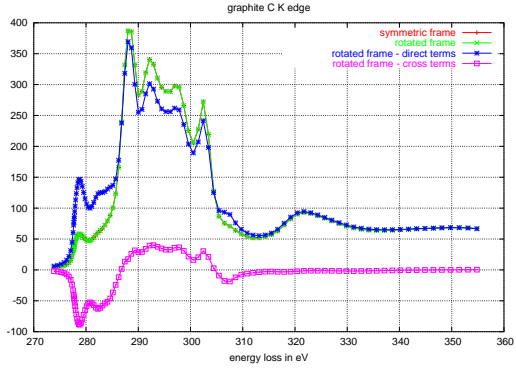}
\caption{C K edge of graphite  calculated in coordinate systems 1 and 2 (see text).  The beam is perpendicular to the graphene sheets, the beam energy
is 300 keV, $\alpha = 10$ mrad, $\beta = 0$ mrad.}
\label{spectrum_rot}
\end{figure}

Finally, Fig.\ \ref{spectrum_rot} shows the resulting ELNES spectrum.  In system 1,
calculation of the direct components of $\sigma$ is sufficient.  To calculate
the same spectrum in system 2, however, the off diagonal components ($yz$
and $zy$ in this example) make a very important contribution.

\subsubsection{\label{magicangle}The Magic Angle}

\begin{figure}
\includegraphics[width=0.4\textwidth]{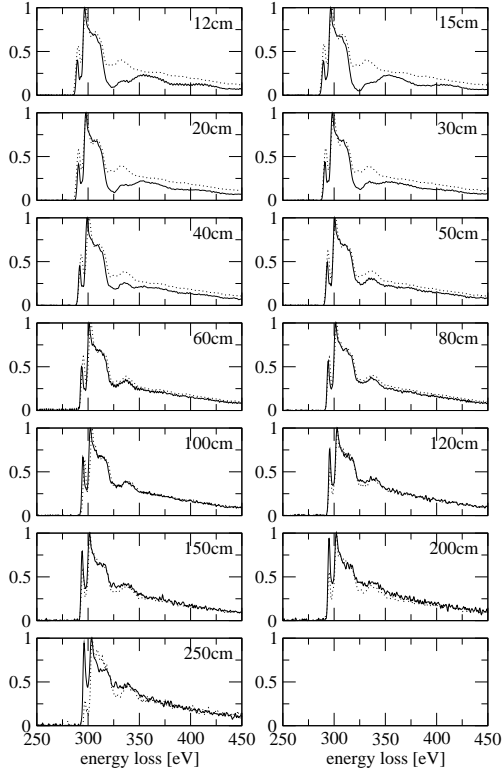}
\caption{C K edge of graphite measured \cite{JoExpt} at 300 keV beam energy for 2 orientations, and different collection angles.  The two orientations overlap between the camera lengths of 80 cm and 100 cm, corresponding to a magic value of the collection angle of about 0.68 mrad.  The convergence angle is close to 0 mrad.}
\label{magic_Jo}
\end{figure}

The magic angle is defined as that value
the collection angle $\beta_m$ of
for which the EELS spectrum is independent of sample to beam
orientation, given a certain convergence angle $\alpha$. 
In the dipole approximation, one can prove easily that such an angle
exists \cite{magichebert} at which the integrals in Eq.\ (\ref{dscssum})
lose their orientation dependence.
The magic angle depends only on beam energy and energy loss (but it is approximately constant over the near edge region) and not on any material property.  The magic angle has played a key role in recent developments of EELS theory \cite{SchattschneiderPRB05} - arguably exactly for that reason.  It is of practical importance to experimentators wishing to eliminate the complications of orientation dependence from their investigations, and turns out to be sensitive to the details of scattering theory.  It is often expressed in units of the "characteristic scattering angle" $\theta_E$, which is the width of the Lorentzian function that approximately gives the DFF as a function of scattering angle 
\begin{equation} \label{thetaE}
\theta_E=\frac{\omega}{E_0}\frac{E_0+m_ec^2}{E_0+2m_ec^2}
\end{equation}
where $E_0$ is the beam energy and $m_e$ is the electron rest mass.
In Fig.\ \ref{magic_Jo}, the magic angle is experimentally found to be $\beta_m = 0.68$ mrad for given experimental conditions.
\cite{JoExpt}  This is close to $ \theta_E = 0.59$ mrad.

\begin{figure}
\includegraphics[width=0.3\textwidth,angle=270]{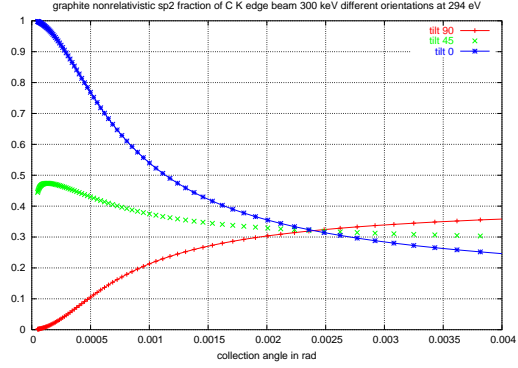}
\caption{$ \frac{\pi}{\sigma} $ ratio (see text) of the graphite C K edge at 10 eV above threshold for a 300 keV beam and three sample to beam orientations.  The magic angle is at the intersection of the three curves.  Nonrelativistic calculation.  The convergence angle is 0 mrad.}
\label{magic_nonrel}
\end{figure}

\begin{figure}
\includegraphics[width=0.3\textwidth,angle=270]{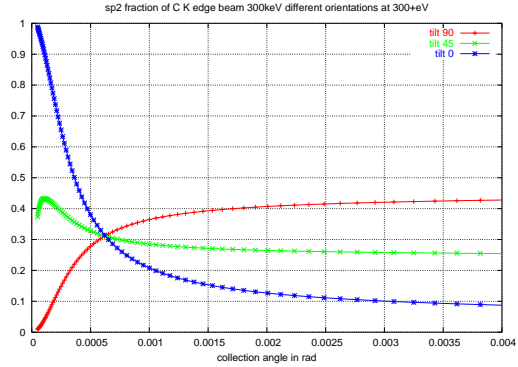}
\caption{$ \frac{\pi}{\sigma} $ ratio (see text) of the graphite C K edge at 10 eV above threshold for a 300 keV beam and three sample to beam orientations.  The magic angle is at the intersection of the three curves.  Relativistic calculation.  The convergence angle is 0 mrad.}
\label{magic_rel}
\end{figure}

We now turn to theoretical calculations using FEFF8.  We calculate spectra at different sample to beam orientations, which we
characterize by a single tilt angle between the electron beam and the crystal $c$-axis.  This tilt corresponds to a rotation
of $ q'_i \,q'_j$ in Eq. \ref{dscssum}.  We could investigate the rotation invariance of differential cross section, but it is more convenient to choose a more sensitive function of the spectrum, and study it as a function of collection angle at fixed energy loss.  At the magic angle, the partial $i=j$ cross sections of Eq. \ref{dscssum} are individually rotation invariant,\cite{magichebert} and therefore we may equivalently study the $ \frac{\pi}{\sigma} $ ratio of the spectrum.
 This function is important as it is related to the $sp^2/sp^3$-ratio which is often used to characterize carbon samples.\cite{jtitantah_sp2}   In symmetric coordinates, it is given by the $\pi$ term or $i=j=z$ term
of Eq.\ (\ref{dscssum}), divided by the $\sigma$ term or $i=j=x$ plus $i=j=y$ term,
\begin{equation} \label{pitosigma}
[\frac {\pi} {\sigma}] := \frac{\sigma_{zz}}{\sigma_{xx}+\sigma_{yy}}
\end{equation}
We calculate this quantity at a fixed energy loss of $E = 294 eV$ as a function of collection angle $\beta$, shown in Fig.\ \ref{magic_nonrel} for a nonrelativistic calculation, and in Fig.\ \ref{magic_rel} for a relativistic
calculation.  Three different sample to beam orientations are shown in each
Figure.  At the magic angle, the spectrum and its $ \frac{\pi}{\sigma} $
ratio are independent of orientation.


\begin{figure}
\includegraphics[width=0.3\textwidth,angle=270]{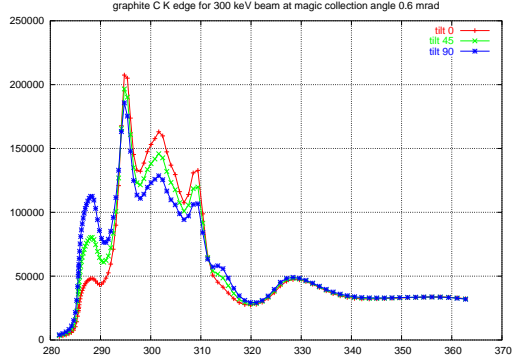}
\caption{C K edge of graphite for 3 orientations, 300 keV beam energy, $\alpha = 0$ mrad, $\beta = 2.4$ mrad).  Relativistic calculation.}
\label{graph_2.4}
\end{figure}

\begin{figure}
\includegraphics[width=0.3\textwidth,angle=270]{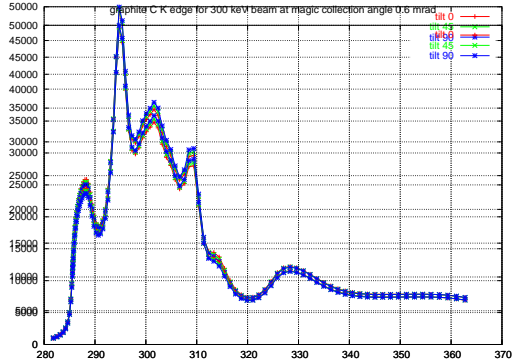}
\caption{C K edge of graphite for 3 orientations, 300 keV beam energy, $\alpha = 0$ mrad, $\beta = 0.6$ mrad).  Relativistic calculation.}
\label{graph_0.6}
\end{figure}

The nonrelativistic simulation in Fig.\ \ref{magic_nonrel} gives $\beta_m = 4 {\theta}_{E}$, as has been reported in the literature for many years \cite{magichebert}, but is inconsistent with experiment.  The relativistic calculation in
Fig.\ \ref{magic_rel} yields a magic angle $\beta_m = 0.65$ mrad, in much better agreement with experimental measurements giving about 0.68 mrad in Fig.\ \ref{magic_Jo}.
The same information is contained in figs. \ref{graph_2.4} and \ref{graph_0.6}, where relativistic calculations of the spectrum are shown at both values of the collection angle.  The nonrelativistic calculation would yield identical spectra at $\beta = 2.4$ mrad, in disagreement with experiment.

Our present results agree very well with calculations reported in \cite{jorissen_UM}, which were calculated using the DFT code WIEN2K.


\section{Conclusions}
We have presented relativistic calculations of electron energy loss spectroscopy using the real space Green's function code
FEFF.  The calculations correctly accounts for ${\bf q}$-dependence and microscope settings such as collection and convergence angle.  We have demonstrated our method on the C $K$ edges of graphite, where we calculate the correct magic angle.

\begin{acknowledgments}
This work is supported in part by the DOE Grant
DE-FG03-97ER45623 (JJR) and was facilitated by
the DOE Computational Materials Science Network.
K. Jorissen gratefully acknowledges financial support
by the F.W.O.-Vlaanderen as Research Assistant of 
the Research Foundation - Flanders.
We wish to thank Z. Levine, C. Hebert, R. Nicholls, D. Lamoen
and P. Schattschneider for comments and suggestions.
\end{acknowledgments}

\appendix

\section{\label{integration}Integrating the cross section over beam and detector}

In this section, we consider the calculation of the differential cross section (Eq.\ (\ref{dscs}) in more detail.
For brevity, we will write $d(E)$ or $d(k')$ for the differential cross section, and $d^2(E,\Omega)$ or $d^2({\bf k},{\bf k'})$ for the double differential cross section, where ${\bf k}$ and ${\bf k'}$ are the wave vectors of the incoming and outgoing beam electron, and ${\bf q}={\bf k}-{\bf k'}$ is their difference.
The detector opening is a circle of radius $\beta$ and the electron beam that hits the sample has a profile $f_\alpha$.
The differential cross section $ d(E) = d(k')$
is then given by
\begin{eqnarray}
\label{app_main}
 d(E) &=& \int_\beta {d\Omega '\,
 \int {d^3k\, d^2({\bf k},{\bf k'}) \,f_\alpha  ({\bf k}) }}\\
  &=& \int { d^3 q\, d^2({\bf q})\, \int_\beta ^{}
  {\,f_\alpha  ({\bf k'} + {\bf q})\,d\Omega '} } \\
  &=& \int { d^3 q \,d^2({\bf q})\, g^{\alpha \beta } (k', {\bf q}) } .
\end{eqnarray}
We have used the fact that $d^2$ is invariant to small rotations of the order of typical scattering angles in EELS (of order mrad), and hence depends only on ${\bf q}$.  A uniform, monochromatic (of fixed energy $k_0^2$), circular beam is described by
\begin{equation}
\label{beam}
f_\alpha  ({\bf k}) = {{I_0 } \over {\pi \alpha ^2 }}\Theta _\alpha  (\theta )\,\,\delta (k - k_0)  
\end{equation}
With this type of beam, the weight $g$ in Eq.\ (\ref{app_main}) is an integral over a constant function,
i.e., a surface.  If the detector aperture is a circle of radius $\beta$, the weight can be interpreted as an overlap of two
circles of radius $\alpha$ and $\beta$ whose centers are separated by the
vector ${\bf q_{\perp}}$.
\begin{equation}
\label{weight}
g^{\alpha \beta } (k',{\bf q}) = {{I_0 } \over {\pi \alpha ^2 }}\,\,\Omega _\beta  \, \cap \,(\Omega _\alpha   - {\bf q_{\perp}})
\end{equation}
which can readily be evaluated using basic algebra.  If collection and convergence angle are interchanged, the shape of the spectrum is conserved, but it is
multiplied by $({\alpha}/{\beta})^2$.
 More complex beam profiles (or detector aperture profiles) could destroy this pseudo-equivalence.

%\bibliography{feffeels}% Produces the bibliography via BibTeX.

\end{document}